\documentclass[epj]{svjour}
\usepackage{graphicx}
\usepackage{bm}
\usepackage{amsmath}
\usepackage{gensymb}
\usepackage{float}
\begin{document}
\title{Resolving solution conformations of the model semi-flexible polyelectrolyte homogalacturonan using molecular dynamics simulations and small-angle x-ray scattering}
\author{Bradley W. Mansel\inst{1,a} \and Amir Hossein Irani\inst{2} \and Timothy M. Ryan\inst{3} \and Duncan J. McGillivray\inst{4,}\inst{5} \and Hsin-Lung Chen\inst{1} \and Martin A. K. Williams\inst{2,}\inst{5,}\inst{6}
}                     
\offprints{bmansel@gmail.com}          
\institute{Department of Chemical Engineering, National Tsing Hua University, Hsinchu 30013, Taiwan \and Institute of Fundamental Sciences, Massey University, Palmerston North 4474, New Zealand \and Australian Synchrotron, Clayton, VIC 3168, Australia \and School of Chemical Sciences, University of Auckland, Private Bag 90219, Auckland, New Zealand \and The MacDiarmid Institute for Advanced Materials and Nanotechnology, Wellington 6140, New Zealand \and Riddet Institute, Massey University, Palmerston North 4474, New Zealand}
\date{Received: date / Revised version: date}
%
\abstract{
The conformation of polyelectrolytes in the solution state has long been of interest in polymer science. Herein we utilize all atom molecular dynamics simulations (MD)  and small-angle x-ray scattering experiments (SAXS) to elucidate the molecular structure of the model polyelectrolyte homogalacturonan. Several degrees of polymerization were studied and in addition partial methylesterification of the otherwise charge-carrying carboxyl groups was used in order to generate samples with varying intra-chain charge distributions. It is shown that at length scales above around 1 nm the conformation of isolated chains has surprisingly little dependence on the charge distribution or the concentration of attendant monovalent salts, reflective of the intrinsic stiffness of the saccharide rings and the dynamical constraints of the glycosidic linkage. Indeed the conformation of isolated chains over all accessible length-scales is well described by the atomic co-ordinates available from fibre diffraction studies. Furthermore, in more concentrated systems it is shown that, after careful analysis of the SAXS data, the form of the inter-particle effects heralded by the emergence of a so-called polyelectrolyte peak, can be extracted, and that this phenomena can be reproduced by multiple chain MD simulations. 
\PACS{
      {PACS-key}{discribing text of that key}   \and
      {PACS-key}{discribing text of that key}
     } 
} 
\maketitle
\section{Introduction}
\label{intro}
Polyelectrolyte solutions typically exhibit complex structures over multiple lengthscales owing to the importance of long-range electrostatic repulsions between chains compared with their uncharged counterparts. These complex structural properties, and similarly complex dynamics, have made polyelectrolytes a fascinating but challenging area of study. The literature on polyelectrolytes is extensive and we point interested readers to the recent review paper by Muthukumar for an extensive overview \cite{muthukumar201750th}. Experimentally polyelectrolytes are often studied using scattering techniques, such as small angle x-ray (SAXS) \cite{ise1979ordered,plevstil1979investigation,ise1980ordered,wang1991small}  or neutron (SANS)  scattering \cite{cotton1976inter,moan1977determination,nierlich1978neutron,nierlich1979small,wang2001small,essafi1999structural,essafi2009hydrophobic,boue2016sans}, due to the suitability of these techniques for accessing structural length-scales in the nanometer regime even for disordered materials. Small-angle scattering measurements carried out on polyelectrolyte solutions under low salt conditions typically reveal the presence of a peak at high scattering vectors together with an increase in scattered intensity at low-q \cite{nierlich1979small}. Below a critical concentration where the chains begin to interact, (often referred to as the overlap concentration), the position of this "polyelectrolyte peak" has been shown to scale with concentration as $C^{1/3}$, in contrast to a $C^{1/2}$ scaling found above this overlap concentration \cite{barrat1996theory,dobrynin2005theory}. The two different regions can be understood from geometric arguments \cite{de1976remarks,barrat1996theory}. Below the overlap concentration the chains occupy individual regions of space often described as blobs. Each chain or blob exerts an electrostatic repulsion on its neighbor, maximizing the distance between them, and giving rise to a correlated distance in the solution, commonly called $\xi$, the correlation length. As the concentration of chains increases the correlation length scales with $\xi \propto C^{1/3}$, in a similar fashion to the distance between points in space as their number increases \cite{de1976remarks,barrat1996theory}. Above the overlap concentration, where the solution can be thought of as a continuous mesh of polymer, the $\xi$ arises due to the distance between "rod-like" regions on neighboring sections of the polymer. Here, the space-filling occurs similar to a random, or Gaussian chain (power law of 2 in 3 dimensions), giving $\xi \propto C^{1/2}$ \cite{de1976remarks,barrat1996theory}.

While small-angle scattering techniques are one of the few experimental methods available to study the solution structure of polyelectrolytes, the modeling of results, in particular in terms of inter-particle interactions can only be decoupled from the overall scattering for certain situations. For centrosymmetric particles the inter-particle effects can be decoupled from those originating from particle size and shape, as long as the particle conformation is known under the identical conditions to those of the more concentrated experiment. For long flexible entities such as polymers, however, the average conformation at length-scales equal to or larger than a persistence length might be expected to change with the concentration of polymer or of added salt (for example \cite{nierlich1985radius}). Consequently, techniques commonly used to screen electrostatic effects for globular particles, such as dilution or salt addition cannot yield a generally applicable form factor that can then be applied under different conditions. In certain situations approximations can be made to try to overcome this issue, but in many cases the decoupling simply cannot be performed. Herein \textit{multi-chain} atomistic molecular dynamics simulations are used in an alternative approach to provide information which can be directly compared to the SAXS experiment. 

Homogalacturonan (HG) is a linear polymer of galacturonic acid residues connected through 1-4 linkages \cite{takagi1979neutron}. Each residue displays one carboxylic acid group, (singly charged above the pKa of around pH 3.5). In addition, a fraction of these groups can be methyl esterified and the existence of well established chemical and enzymatic tools means that both the charge density and distribution can be systematically modified \cite{kent2016structure}. This ability to transform the same polymeric backbone from a strong polyelectrolyte with a linear charge density of $~ 2e^-/nm$ to a moderately, or even uncharged polymer, makes it an interesting subject for the controlled study of polyelectrolyte effects.

As with many other biopolymers homogalacturonan is expected to display a complicated inter-chain potential. While at large length-scales electrostatic interactions are expected to dominate, at smaller length-scales inter-chain hydrogen bonding and hydrophobic effects could have significant influences. In fact, it is well known that under certain conditions gelation can occur (even in the absence of the commonly exploited divalent crosslinkers) \cite{gilsenan2000thermally,mansel2015zooming} as a result of these interactions. Interestingly, the propensity for such gelation has been shown to depend on the precise helical conformation of the chains, with a transition from a predominantly 2,1 to 3,1 helix inducing significant inter-chain attraction in the form of hydrogen bonding. Previous experimental and computational studies have shown that in the solution state at pH values near 7 the chain conformation is primarily 2,1, but with a certain percentage of the monomers in a 3,1 conformation \cite{cros1996solution,braccini1999conformational,jarvis1995chain,noto2005role}. It must be acknowledged that, in-light of these complex structural properties the inter-chain potential for homogalacturon (as well as similar carbohydrates, such as alginate), is likely to be highly complex.

Previous small-angle scattering studies on the structure of polyelectrolytes in solution have primarily focused on DNA and polystyrene sulfonate (PSS) which have been considered model semi-flexible and flexible polyelectrolytes, respectively. As has been discussed previously, extracting inter-particle effects from polyelectrolytes can only be performed under certain situations. For DNA, at length-scales much shorter than a persistence length the structure can be approximated as a rod, described simply by a length and radius, and this has been utilized to extract inter-particle interactions and obtain an effective structure factor \cite{van1998structure}. In another study oligomeric double stranded DNA (25 basepairs) was studied under different low salt conditions \cite{qiu2006measuring} and a decoupling approximation was utilized to account for the particle anisotropy \cite{kotlarchyk1983analysis,greene2016critical}. For PSS, early neutron scattering experiments on mixtures of deuterated and non-deuterated PSS have been performed \cite{nierlich1979small,jannink1986structure,nierlich1985radius,nierlich1985characteristic}. By contrast matching the deuterated chains using a mixture of water and deuterium oxide the single chain form-factor could then be obtained under essentially identical conditions to those of the semi-dilute solutions. From this approach information about inter-particle interactions could be experimentally obtained in the form of a discrete correlation function. Other work on alginate approached the modeling of SAXS data using a number of analytical models \cite{josef2012conformation}. Alginate shares many similarities with homogalacturonan and while these methods could be applied to our experimental data many assumptions must be made (the persistence length calculated ranged to values below 1 nm, which differs significantly from the conclusions of the MD simulations performed during our work).

Here, we first present SAXS data from short HG oligomers that can be modeled using MD in a reasonable time, and second we investigate longer chains (commonly referred to as pectins) which hold significant interest for industrial purposes. Multiple homogalacturonan chains (of around 25 monomers long) could be modeled using modern super computer facilities in a reasonable time period. \textit{These simulations provided us with the atomic coordinates for both the form of the scatterers and their spatial arrangement.} It is shown that the single chain structure can be reproduced accurately in-silico and subsequently, that with this approach we can elucidate many properties that are required to accurately analyze the SAXS data and are often difficult or impossible to obtain from the experiment alone. For instance, we can verify whether increasing the concentration has a significant effect on the single chain structure, a vital piece of information when investigating inter-particle effects. The helical conformation of chains under different conditions and the effects of added salt can also be easily obtained. These effects are difficult to extract from experiment alone, as changing, for example, the ionic conditions can have subtle effects on the solution structure as well as introducing further scattering from introduced ions. The low concentrations required to have isolated scatterers means a low signal to noise ratio, which once more makes the experiments challenging. Here we directly compare inter-chain effects from SAXS and MD, and show that the multi-chain MD simulations capture the emergence of the polyelectrolyte peak and its attendant scaling.

\section{\label{sec:methods}Methods and materials}
\subsection{\label{sec:sample_prep}Homogalacturonan solution preparation}

\paragraph{Homogalacturonan Oligomers} with degrees of polymerisation (DP) of 6 and 10 were homemade as described in the literature \cite{williams2002analysis}. Briefly, polygalacturonic acid was subject to a partial digestion with an endo-acting enzyme, endo-PG II, and the digest products were separated from each other using high performance anion exchange chromatography (HPAEC). Fractions containing the polyelectrolyte of interest were pooled from several separations to obtain reasonable quantities. These samples will be referred to as $DP6$ and $DP10$. An oligomeric sample containing DPs between 25-50 was purchased from Elicityl (http://www.elicityl-oligotech.com) and will be referred to as $DP25$. In total three oligomeric polymers were studied. 

\paragraph{Pectin Samples} consisting predominantly of homogalacturonan (85\% galacturonic acid content), with $DP\approx 500$, were obtained from CP Kelco. Previously characterized samples (of differing degrees of methylesterification, as given), and charge distributions (denoted random, R, or blockwise, B) were examined \cite{irani2017molecular}. A single blocky pure HG was also made by removing the homogalacturonan sections $DP\approx 100$ from a commercial sample with a mild acid treatment, highly methylesterifying and then treating with a processive enzyme as previously described \cite{williams2010extracting}; this sample is denoted as $HG40$.

\paragraph{Ion Control} for all SAXS measurements consisted of first running sample solutions through an ion-exchange column (Amberlite) to remove any residual ions, and render the galacturonans in their acid form. These were then freeze dried and redissolved at the required concentrations. The pH of solutions was modified to $p\textnormal{H}=7$ using a calibrated pH meter and 0.1 M NaOH to give pH-controlled samples with the lowest salt concentrations possible. For samples requiring an additional concentration of monovalent ions, the required amount of dry NaCl was added. Finally all solutions were filtered through 0.2 micron syringe filters to remove any contaminants. Static and dynamic light scattering showed all solutions to be dust free after the procedure.
 
\subsection{\label{sec:method_saxs}Small-angle x-ray scattering}

\paragraph{SAXS experiments} were performed on the small and wide angle x-ray scattering beamline at the Australian Synchrotron \cite{kirby2013low}. Experiments were performed over two beam-time allocations with identical configurations. Samples were held in 1.5 mm path-length quartz capillaries and filled using an automated flow cell allowing measurement of the sample and background in the same positions on the capillary, to give accurate background subtraction. Sample irradiation, with an energy of 15 keV, was performed for multiple 1 second intervals with the sample flowing at 5 $\mu$L per second. A Pilatus 1M detector located $ \approx 0.7 m$ from the sample, (slightly different in experiments carried out during the two different beamtime allocations), collected the scattered radiation. For each set of experiments  the sample to detector distance was calibrated using an AgBeh standard. The intensity of scattered radiation was normalized to the incident beam and put on an absolute scale using a glassy carbon standard. Corresponding buffer background experiments were performed by adding the same amount of ions, as per the samples, including any added acid or base, to ultra pure (18.2 $M \Omega$) water from the same source used to make samples. These buffer experiments were subtracted from the sample measurements using scatterBrain, the standard beamline software. For this study the magnitude of the scattering vector is defined as $q=|\mathbf{q}|= 4\pi \lambda^{-1} \sin(\theta/2)$ where $\theta$ is the angle between the incident beam and scattered radiation.

\subsection{\label{theory}SAXS analysis}
For oligomer samples DP6 and DP10, while we cannot obtain the structure factor from SAXS, defined as the Fourier transform of the center-of-mass (COM) inter-particle correlation function, we can instead obtain the monomer-monomer inter-particle correlation function from the scattering data. Crucially, this can also be directly calculated from MD simulation and compared to our experimental SAXS results. \textit{This is only possible for cases where the isolated particle conformation does not change with inter-particle effects, and for these oligomers that was verified using the MD simulation}.

A similar approach to that first described by Williams et. al. is utilized here \cite{williams1979polyelectrolyte} and provides a general approach for analyzing interacting polymer systems \cite{jannink1986structure,krakoviack2002relating,van1998structure,nierlich1985characteristic,nierlich1985radius,nishida2002added}.  Excluding contributions from ions and water, the intensity of scattered x-rays is related to the scattering length density (SLD) difference between water and galacturonic acid $\Delta \rho$, the volume of scatterer $V$, and the monomer-monomer spatial correlation function $H(\mathbf{q})$ by:
\begin{equation}
I(q)\propto \Delta \rho^2 V^2 H(\mathbf{q})
\end{equation} 

\begin{equation}
H(\mathbf{q}) = \sum_{i,j} \sum_{\alpha, \beta} \exp[iq(\mathbf{r}_i^\alpha-\mathbf{r}_j^\beta)]
\end{equation}
where $\alpha$ and $\beta$ are chain indices and $i$ and $j$ are monomer indices along each chain. $H(q)$ can be separated into two parts, one related to a sum of intra-molecular correlations and the other inter-molecular correlations:
\begin{equation}
	H(\mathbf{q}) = \sum_{i,j} \sum_{\alpha=\beta} \exp[iq(\mathbf{r}_i^\alpha-\mathbf{r}_j^\beta)] + \sum_{i,j} \sum_{\alpha \neq \beta} \exp[iq(\mathbf{r}_i^\alpha-\mathbf{r}_j^\beta)]
\end{equation}
which we will refer to as the self $H(q)^s$ and distinct $H(q)^d$ components. For SAXS, where all the monomers have the same scattering length density, (compared to SANS where the SLD is often varied to provide contrast matching through selective deuteration) $H(q) = H(q)^s + H(q)^d$. The self component is related to the form factor, $P(q)$, or normalized self correlation function, through $P(q)=H(q)^s_n=H(q)^sN_M/N $, where $N$ represents the number of monomers per unit volume and $N_M$ the number of monomers per chain, while $H(q)^d$ is related to the monomer-monomer radial distribution function $g_{mm}$ through:
\begin{equation}
H(\mathbf{q})^d = \int d^3r \exp[{i\mathbf{qr}}](g_{mm}(\mathbf{r})-1)
\end{equation}
where
\begin{equation}
g_{mm}(\mathbf{r})=\frac{1}{N^2} \sum_{\alpha \neq \beta} \sum_{i,j} \langle \delta(\mathbf{r}-\mathbf{r}_{i}^\alpha +\mathbf{r}_{j}^\beta ) \rangle
\end{equation}
and can be calculated from the monomer COM positions obtained from the MD simulations. From the SAXS experiments $g_{mm}$ can be obtained from the discrete Fourier inversion \cite{skinner2016structure,narten1976diffraction,okhulkov1994x} of the normalized distinct correlation function $H(q)^d_n=H(q)^d/N^2$ through:

\begin{equation}
g_{mm}(r)=1+\frac{dq}{2\pi^2 \rho r}\sum_{q_{min}}^{q_{max}} qH(q)^d_n \sin(qr) \frac{\sin (q\Delta r)}{q\Delta r}
\label{eqn:Fourier_inversion}
\end{equation}

where $\rho$ is the number of chains per \AA$^3$ ($1.189 \times 10^{-5}$ and $7.134 \times 10^{-6}$ \AA$^{-3}$ for DP6 and DP10, respectively), and $dq$ is the $q$ spacing between successive data points (equal to 0.018 \AA$^{-1}$ for these experiments). The final sine function accounts for smoothing the data over a length scale $\Delta r$ and was set to $4$ \AA.

The correlation length, $\xi$, was extracted by fitting the intensity data to a commonly used empirical model, known as the broad peak model, consisting of a low-q power-law and Lorentzian function:
\begin{equation}
I(q) = \frac{D}{q^n}+\frac{E}{1+(\lvert q-q_0 \rvert \zeta)^m}
\end{equation}
where the correlation length is defined by $\xi = 2\pi/q_0$, $D$ and $E$ are scaling constants, $n$ is the power law exponent, $m$ is the Lorentzian exponent and $\zeta$ the screening length.

\subsection{\label{sec:method_MD}Molecular dynamics simulations}
\paragraph{Molecular Dynamics (MD) simulations} were performed using GROMACS \cite{Berendsen199543}. Simulations were carried out based on the GLYCAM force field \cite{woods1995molecular} with the partial charges of atoms assigned according to previous DFT calculations \cite{irani2017molecular}. SPC/E water molecules \cite{berendsen1987missing} were used. The temperature was maintained at 298 K using a Berendsen thermostat \cite{berendsen1984molecular}, all bonds (even heavy atom-H bonds) were kept constrained, and the pressure was kept fixed at 1.0 bar using the Parrinello-Rahman algorithm \cite{ParrinelloRalgorithm}. Periodic boundary conditions were applied and the integration step was set to 2 fs. The Particle Mesh Ewald method \cite{darden1993particle} was used in order to treat the long-range electrostatics. The simulations were performed using supercomputing facilities available through NeSI. Between 20 and 250 ns of simulation could be carried out per day depending on the size of the simulation box. The box size was chosen to ensure at least a 1 nm distance between any chain and its image in the neighboring box formed by periodic boundaries. The size of the box was between 12 and 20 nm  when containing between 3 and 50 chains respectively depending on the required concentration and available computing resources, typically necessitating the order of 10s to 100s of thousands of water molecules. For the DP6 and DP10 simulations, which were compared to experiment, 20 and 50 chains were contained in a cubes of length 11.6 and 18.6 nm, respectively.  The initially generated configurations of the oligogalacturonides were equilibrated using interaction parameters from the GLYCAM force field  \cite{woods1995molecular} and the PyMOL builder module \cite{PyMOL} was used to perform methyl-esterification of selected galacturonic acid residues. MD visualization was performed using VMD \cite{humphrey1996vmd}. The whole system was neutralised by Na$^+$ cations. 200 ps of simulation was performed first for the NVT and subsequently the NPT ensemble prior to the start of the simulation proper, which was sufficient to achieve equilibrium temperature, pressure and density.

To directly compare the structure of isolated chains from SAXS and MD simulations, Crysol, from the ATSAS software package, was utilized \cite{svergun1995crysol}.The software converts the real-space atomic coordinates obtained from simulations into radially-integrated reciprocal space data that can be directly compared with SAXS data. 

\section{Results and discussion}
\label{results}
\subsection{\label{sec:isolated_chain_scattering} Isolated chains}

\begin{figure}
\begin{center}
	\includegraphics[width=0.85\columnwidth]{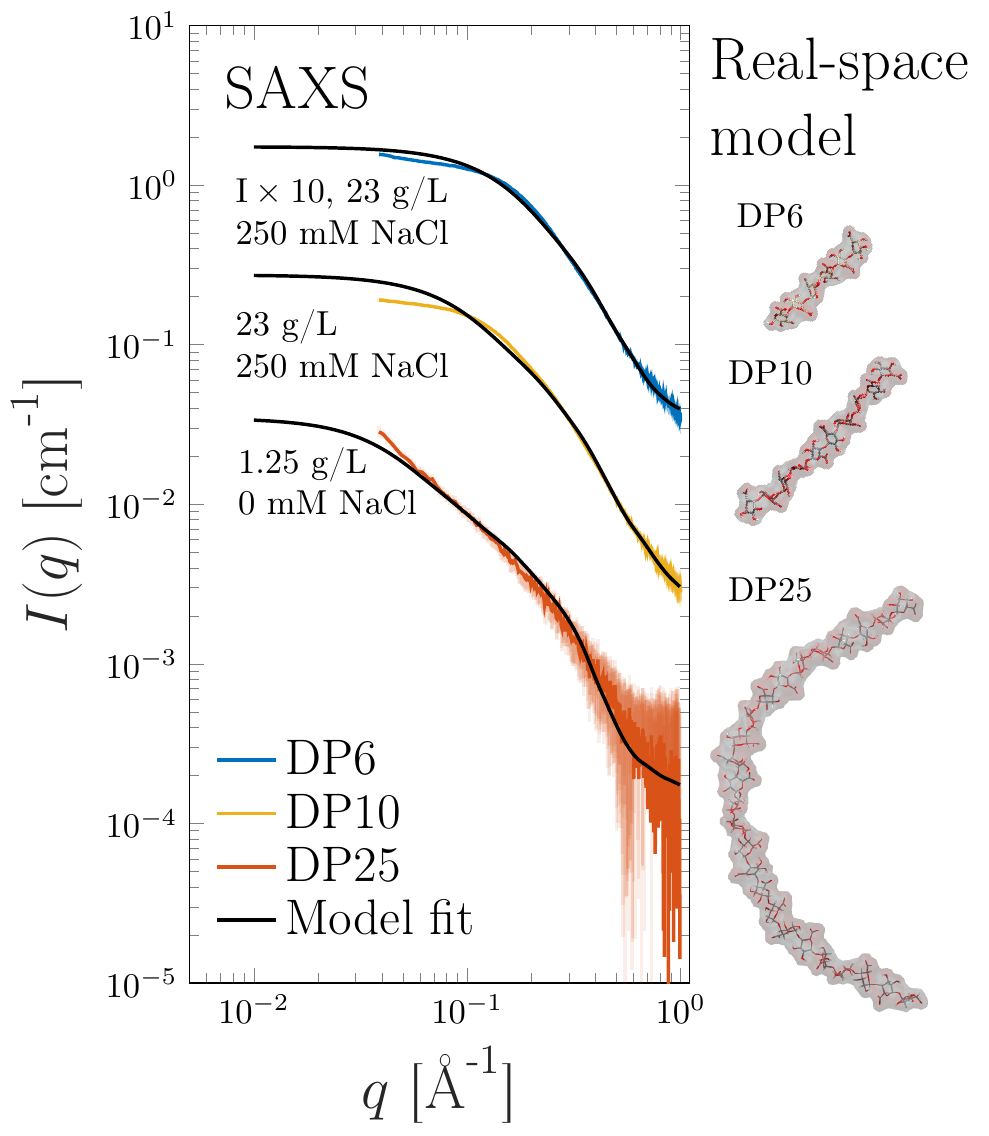}    
	\caption{SAXS data from homogalacturonan with differing degrees of polymerization (DP). Conditions (low concentration or high ionic strength) were selected in an attempt to minimize the structure factor. Black lines show  the calculated scattering from the atomistic models shown on the right.}
	\label{fig:2_saxs_crysol}
    \end{center}
\end{figure}

\begin{figure}
\centering
	\includegraphics[width=\columnwidth]{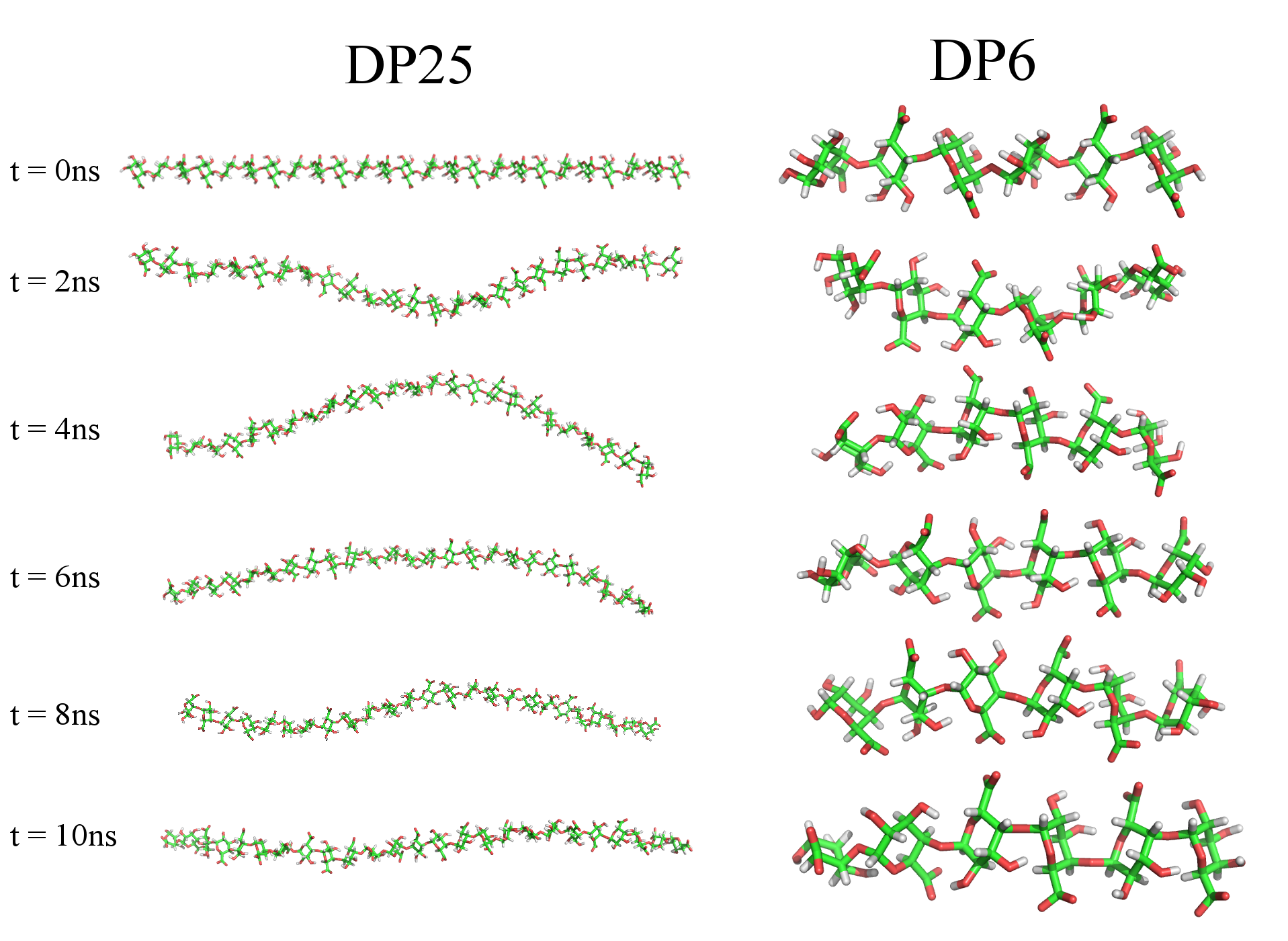}
	\caption{Chain conformation for DP25 and DP6 at different times during the simulation. At length-scales comparable to 6 monomers the chains display little flexibility, while over a few tens of monomers the flexibility of the glycosidic linkages can be observed. This displays the length-scales over-which flexibility becomes important in modeling the SAXS data.}
	\label{fig:sim_chain_conformation}
\end{figure}

A comparison between experimentally measured scattering patterns obtained from SAXS (at low concentration or where significant amounts of salt were added) and those calculated from the output of MD simulations is shown in figure \ref{fig:2_saxs_crysol}. It can be seen that the MD model is representative of the SAXS data for $DP25$ across the q-range. Deviations at low-q, for the $DP6$ and $DP10$ samples, can be attributed to inter-particle effects. (Attempts at retaining scattering intensity while removing structure factor effects by adding 250 mM salt to the system were not completely successful). From MD analysis,  the conformations of DP6 and DP10 were found to be rod-like, (see figure \ref{fig:sim_chain_conformation}). For the $DP25$ sample it was found that the agreement between experiment and simulation relied on the flexibility of the polymer to some degree. To account for the flexibility and dynamics,  50 structures from the ensemble were output (for example figure \ref{fig:2_saxs_crysol}) and used to produce SAXS profiles using Crysol, which were then averaged.  

It can be seen that the chain generally conforms to an extended profile, figure \ref{fig:sim_chain_conformation}, during the entire simulation. Previous simulation studies have found that the polymer backbone has 2-3 energetically favorable local conformations which can give rise to different persistence lengths. Indeed, Noto \cite{noto2005role} found that for charged polygalacturonic acid the asymptotic persistence length fluctuated significantly during the 3 ns simulation giving a persistence length of $300 \pm 50$ \AA $\space$ with the large uncertainty related to the large fluctuations in persistence length during the simulation (related to fluctuations of some parts of the chain between 2,1 and 3,1 helices). This was also found during our simulation.  We conclude that within the current limitations of MD simulation and SAXS analysis our simulation and experimental results agree reasonably well for isolated chains. 

\subsection{Environmental effects on the chain flexibility}
It might be expected that for a semi-flexible polyelectrolyte the electrostatic persistence length as proposed by Odijk \cite{odijk1977polyelectrolytes}, and Skolnick and Fixman \cite{skolnick1977electrostatic}, and commonly referred to as the OSF model should be relevant. The theory is based on the idea that the persistence length of polyelectrolytes is governed by the effects of stiffening from the repulsion of charged groups along the chain in addition to the intrinsic persistence length; specifically: $l_{p, eff} = L_{pi} +l_{pe} $ where $l_{p, eff}$ represents the effective persistence length (that measured in solution), $L_{pi}$ the intrinsic persistence length and $l_{pe} $ the electrostatic persistence length. This electrostatic persistence length is given to first order by \cite{odijk1977polyelectrolytes} $l_{pe} = l_B\kappa^{-2}A^{-2}/4 $ where $l_B$ is the Bjerrum length, $\approx 7${\AA} \cite{muthukumar201750th} in water at room temperature, $\kappa^{-1}$ is the Debye screening length, and $A$ is defined as the contour distance between adjacent "beads". Even taking $A$ to be the smallest possible value (the length of a monomer) in order to maximize $l_{pe}$, we find the electrostatic persistence length is only $\approx5$ {\AA}, approximately the size of one monomer. This is due to the relatively large ionic concentration that arises from the significant base that is required to change the pH of the sample to 7, where all the galacturonic acid groups can safely be assumed to carry a charge. The 130 mM of monovalent salt arising from changing the pH of the 23 g/L solutions produces a Debye screening length of $\approx8$ {\AA}, so that the electrostatic persistence length model does not need to be taken into account for describing the solution conformation of these homogalacturonan samples.

This was further explored by performing the MD simulation in the presence of 100 mM additional NaCl where, consistent with our contention little difference was observed in the flexibility of chains. This also confirms that screening inter-particle effects using salt, in this specific case, has little influence on the single chain structure. 

For homogalacturonan, the effect of charge on the persistence length has been shown to be governed in a more hierarchical manner, with local conformational changes along the chain altering the disposition of the saccharide rings\cite{noto2005role}. This is a significantly different mechanism to the electrostatic persistence length model. Experimentally and computationally two distinct conformational structures of the chains have been found to represent the most populated configurations of the chain: that of 2-fold or 3-fold helices \cite{noto2005role,chandrasekaran1999x,braccini1999conformational,morris1982conformations,gilsenan2000thermally,manunza1997molecular,manunza1998interaction}, with the dominant structure depending on the charge state of the polymer. The simulations reported here were undertaken in order to model the experimental data and thus the in-silico chains were in the fully charged state, and were consequently found (as previously reported) to indeed be predominantly in the 2-fold helical arrangement (albeit with fluctuations as described above). 

\subsection{\label{sec:Interacting chains}Interacting oligomers}

\begin{figure}
\centering
	\includegraphics[width=\columnwidth]{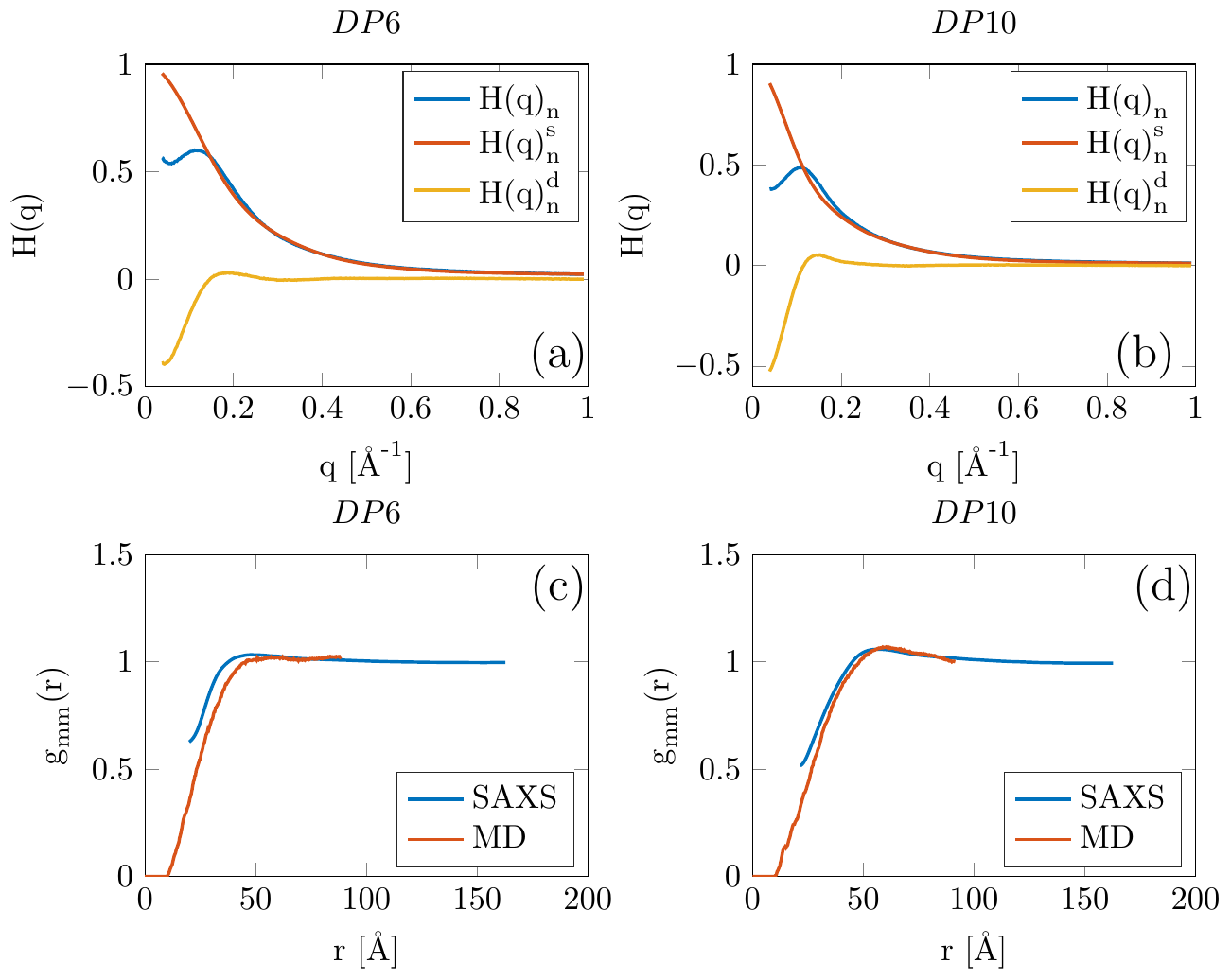}
	\caption{Figures (a) and (b) show normalized total ($\textnormal{H(}q\textnormal{)}_\textnormal{n}$), self ($\textnormal{H(}q\textnormal{)}^\textnormal{s}_\textnormal{n}$) and distinct ($\textnormal{H(}q\textnormal{)}^\textnormal{d}_\textnormal{n}$) correlation functions for DP6 and DP10. Figure (c) and (d) shows a comparison between $g_{mm}(r)$ for DP6 and DP10 obtained with MD and SAXS at a concentration of 23 g/L.}
	\label{fig:4_SAXS_SIMS_dps}
\end{figure}

For 23 g/L DP6 and DP10 samples significant inter-particle interactions could be measured in the SAXS experiments after reducing the concentration of monovalent salt added to the samples to the minimum possible, (while still achieving the pre-requisite pH change)  (figure \ref{fig:4_SAXS_SIMS_dps}(a)). Here, in keeping within the formalisms outlined in section \ref{sec:methods}  both the normalized total, self and distinct correlation functions are plotted. \textbf{The self part} of the correlation function, also known as form factor, was calculated from the molecular coordinates output from the \textbf{MD} and was scaled in intensity by setting the value at zero angle to unity. The \textit{normalized total correlation function}, $\textnormal{H(}q\textnormal{)}_\textnormal{n}$, was obtained by scaling the \textit{observed scattering} by a factor so that $\textnormal{H(}q\textnormal{)}^\textnormal{s}_\textnormal{n}/\textnormal{H(}q\textnormal{)}_\textnormal{n}=1$ in the high-q region. This can be justified as in our experiments we obtain data to sufficiently high $q$ ($q=1$ \AA$^{-1}$) that the scattering is completely dominated by $\textnormal{H(}q\textnormal{)}^\textnormal{s}$. At such small length-scales (close to the monomer-monomer distance) intra-chain correlations are significantly more likely than inter-chain.

From figure \ref{fig:4_SAXS_SIMS_dps}(a) and \ref{fig:4_SAXS_SIMS_dps}(b) it can be seen that $\textnormal{H(}q\textnormal{)}^\textnormal{s}_\textnormal{n}$ and $\textnormal{H(}q\textnormal{)}_\textnormal{n}$ are indeed nearly identical above $q=0.4$ \AA$^{-1}$, while below these values significant inter-particle effects can be observed in $\textnormal{H(}q\textnormal{)}_\textnormal{n}$. $\textnormal{H(}q\textnormal{)}^\textnormal{d}_\textnormal{n}$ was simply obtained using $\textnormal{H(}q\textnormal{)}^\textnormal{d}_\textnormal{n}=\textnormal{H(}q\textnormal{)}_\textnormal{n}-\textnormal{H(}q\textnormal{)}^\textnormal{s}_\textnormal{n}$. Subsequently $g_{mm}(r)$ is calculated from the scattering data by performing a Fourier inversion (equation \ref{eqn:Fourier_inversion}), with the data for the DP6 and DP10 samples shown in figures \ref{fig:4_SAXS_SIMS_dps}(c) and \ref{fig:4_SAXS_SIMS_dps}(d) respectively. $g_{mm}(r)$ could only be calculated to 20 \AA  $\space$ due to artifacts at smaller length-scales resulting from the Fourier inversion, but it can be seen that there is a reasonable agreement between the SAXS measurement and the multi-chain MD. 

$g_{mm}(r)$ describes the probability of finding a monomer from another chain at a distance $r$ from a reference residue. It can be considered to be similar to $g(r)$, although the function is smeared over the form of the chains compared to $g(r)$ which describes correlations between the centers of masses of chains. If $g_{mm}(r)$ is less than one it means that there is a lower chance of finding a monomer belonging to another chain (which the reference monomer does not belong to) than solvent at that distance, while if it is greater than one then a higher chance exists to find the monomer of another chain than solvent.
For these samples we see that at short distances there is a small chance of finding different chain monomers, which increases to a maximum at around $40$ \AA $\space$ for the DP6 sample and $50$ \AA $\space$ for the DP10 sample (\ref{fig:4_SAXS_SIMS_dps}(c) and \ref{fig:4_SAXS_SIMS_dps}(d)).

In these two samples there are the same number of monomers (each carrying a negative charge), except arranged in different numbers of linear groups of six or ten residues comprising the chains. The DP10 sample contains a lower number density of chains compared to DP6, but with each more charged, so that it is expected that the peak in $g_{mm}(r)$ is at larger distances, as found. At length-scales $<25$ \AA $\space$ we see that $g_{mm}(r)$ still has a small ($g_{mm}(r)<0.5$) value indicating a small likelihood that monomers from different chains will be located so close to each other. It appears likely then that some attraction exists at short length-scales, although quantifying these interactions is difficult due to the anisotropic nature of the particles and potentials.

\subsection{Interacting polymers}

\begin{figure*}[h!]
	\includegraphics[width=0.9\textwidth]{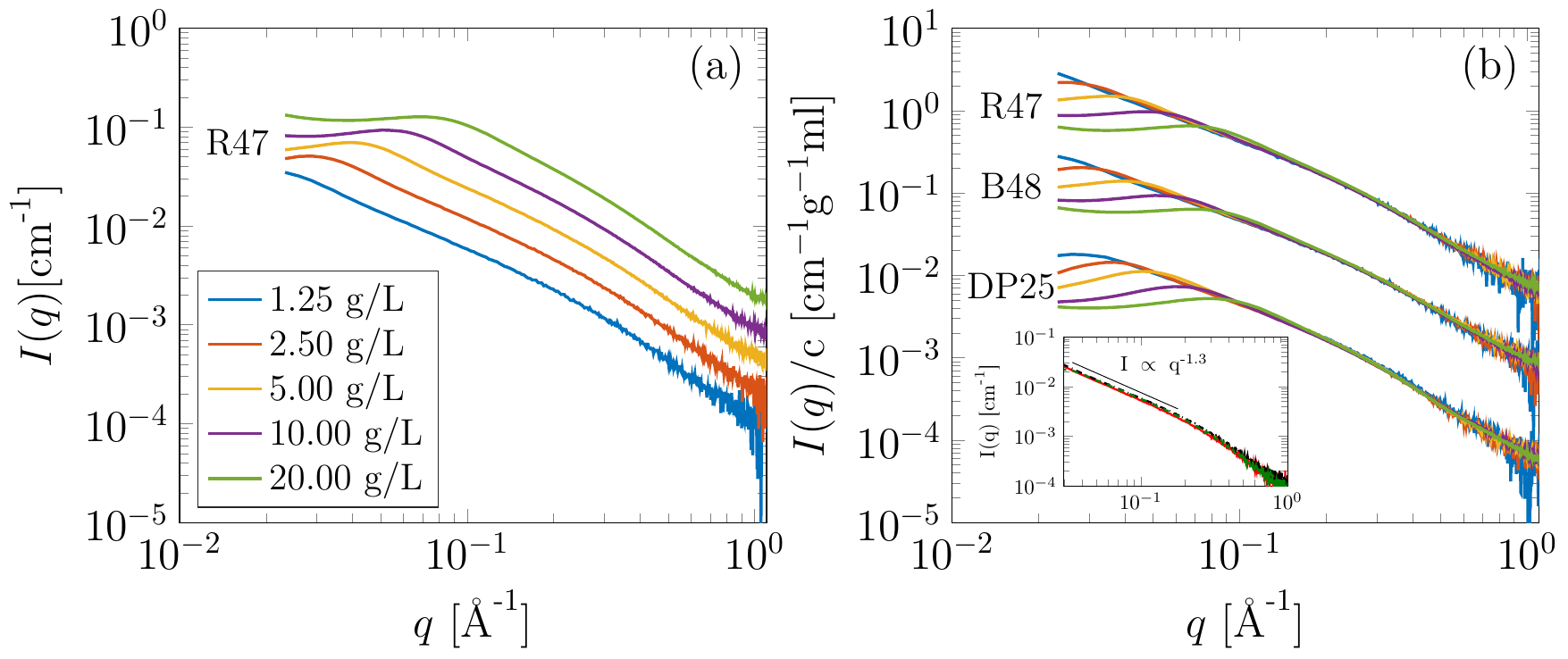}
	\caption{(a) SAXS data from homogalacturonan with a random charge distribution, R47, at different concentrations, 1.25, 2.50, 5.00, 10.00 and 20.00 g/L. (b) Concentration normalized SAXS data from homogalacturonan with three different charge distributions showing the influence of electrostatic inter-chain repulsion at higher concentrations. The line color represents the concentration, as shown in the legend of figure (a). Data for R47 and DP25 has been scaled by a factor of 10 and 0.1, respectively.  The insert in (b) shows the 1.25 g/L for the three samples on an absolute scale to highlight similarities in the scattering at q larger than the correlation peak (if present).  }
	\label{fig:3_saxs_B48_R47}
\end{figure*}

Figure \ref{fig:3_saxs_B48_R47} shows SAXS results for samples with a considerably larger degree of polymerization of approximately 500 residues, corresponding to an approximate contour length of 250 nm. (In detail these samples contain something of the order of 1\% rhamnose residues in the backbone and might conventionally be referred to as pectin, containing some small amount of substituted rhamnogalacturonan in addition to HG). In addition, and in contrast to the oligomeric samples predominantly described thus far, these samples have varying degrees of methylesterification that has been used to control the amount and patterning of the charges carried by the galacturonic acid residues. Figure \ref{fig:3_saxs_B48_R47}(a) shows the data on an absolute intensity scale for R47 (R=randomly distributed charged groups \cite{irani2017molecular}, 47=47\% of the galacturonic acid residues are methylesterified and cannot carry a charge), as a function of concentration, in the absence of added monovalent salt. It can be seen that at 0.125 wt\% no inter-particle effects can be observed whereupon by increasing the concentration a prominent inter-particle effect occurs. Figure \ref{fig:3_saxs_B48_R47}(b) shows the same data, normalized by concentration, together with results from B48, (B=blockwise distributed charged groups \cite{irani2017molecular}, 48=48\% of the galacturonic acid residues are methylesterified), and DP25 (short chains of between 25 and 50 residues long consisting of entirely charged homogalacturonic acid with no methylesterification). At low concentrations and / or q above $q=0.1$ {\AA}$^{-1}$ the three different samples produce highly similar scattering. For the B48 and R47 samples at 1.25 g/ml concentration no inter-particle effects can be observed, while for the highly charged DP25 sample a small structure factor is still evident even at this low concentration (for these measurements a slightly different sample to detector distance was utilized when performing the concentration series which gives access to lower $q$ than is presented in figure \ref{fig:2_saxs_crysol}). For these longer chains as the concentration increases it is highly likely that the form of the individual particles changes as well as the inter-particle effects and as such, to the authors knowledge, no quantitative methods to analyze the data exists. While for the DP6 and DP10 calculations MD simulations could be used in concert with the experimental data to investigate the intra and inter chain contributions to the scattering directly, multi-chain MD of many 500 residue chains was too computationally expensive for this study to be undertaken.

\subsubsection{Invariance of isolated chain scattering with fine structure}
It can be seen from the inset in figure \ref{fig:3_saxs_B48_R47}(b), which shows the three samples at 1.25 g/L on an absolute scale, that above $q=0.04$ \AA$^{-1}$ little structural difference exists between the samples. In the lowest concentration (single chain) scattering little difference can be observed between three samples with 2 different DPs and three different charge distributions.  To extract a persistence length from SAXS data the worm-like chain model is typically used to describe the data. It predicts power law scattering with slope -1 that transitions to a self avoiding walk (SAW) with slope -5/3 (power law of -5/3) at lower q. Interestingly, in our data neither a power-law of -1 nor -5/3 is observed, but instead an intermediate value of -1.3. Fitting of a power law of form $I(q) \propto q^{-n}$, over $0.04 \leq q \leq 0.23$ \AA$^{-1}$ produced values of $n = 1.31 \pm 0.02, \space 1.34 \pm 0.02, \space 1.31 \pm 0.01$ with $\chi^2/N_{points}= 0.8321, \space 0.556, \space 0.2178$ for the KB48, KR47 and DP25 samples, respectively. The number of data points in the fitted region, $N_{points}=69$. While the lowest q we access here is potentially not small enough to access the SAW walk scattering, at the q values we did investigate power law scattering with an exponent of -1 is expected. When the authors tried fitting the scattering data with a worm-like chain model, using the commonly utilized SASView software, the  -1.3 power-law observed experimentally could not be produced from the model. This shows that one distinct structural length-scale for the persistence length does not exist, making the application of the worm-like chain model questionable for the analysis of SAXS data from homogalacturonan. Interestingly this seemingly controversial finding is consistent with the MD simulations where, as described earlier, a  distribution of rod like (2,1 helical sections) and more flexible (3,1 regions), can be observed within individual chains and across the ensemble. As such, a distribution of persistence lengths is expected, so that the lengthscale at which individual chains exhibit the -1 or -5/3 scaling exponent is different, explaining the averaging of the behavior observed. 

\subsubsection{Invariance of correlation peak position with fine structure}
Initially it might be expected that the appearance of regions of large $local$ charge density in the samples with a blockwise charge distribution would drive more significant inter-chain repulsion. However, a previous study carried out on R47 and B48 \cite{irani2017molecular} investigating the electrophoretic transport of these polyelectrolytes has shown that significant counter-ion condensation exists for the B48 sample in contrast to the  randomly distributed charges of R47, with the maximum charge renormalization on galacturonic blocks reached when around  30 \% of the residues bind ions tightly enough to essentially appear uncharged.  Consequently, even though the over-all chain-averaged charge density of the blocky sample is less that the random analogue because of condensation there will still be blocks of contiguous residues with local charge densities equivalent to having chains with 70 \% charge (compared with the 53\% charge on the random sample), and these ion-decorated but still highly charged local regions will still give rise to inter-chain repulsion. These highly charged regions are however distributed amongst less charged regions and are therefore effectively at a lower concentration, which mitigates the effect to some extent, and explains why there is little difference between the behavior exhibited by the different fine structures. This invariance of the polyelectrolyte peak position and shape with fine structure (as a result of charge renormalization) has previously been observed in another copolymer system \cite{essafi1999structural}.

\subsection{\label{sec:scaling_laws} Scaling laws}

\begin{figure}[h!]
	\includegraphics[width=\columnwidth]{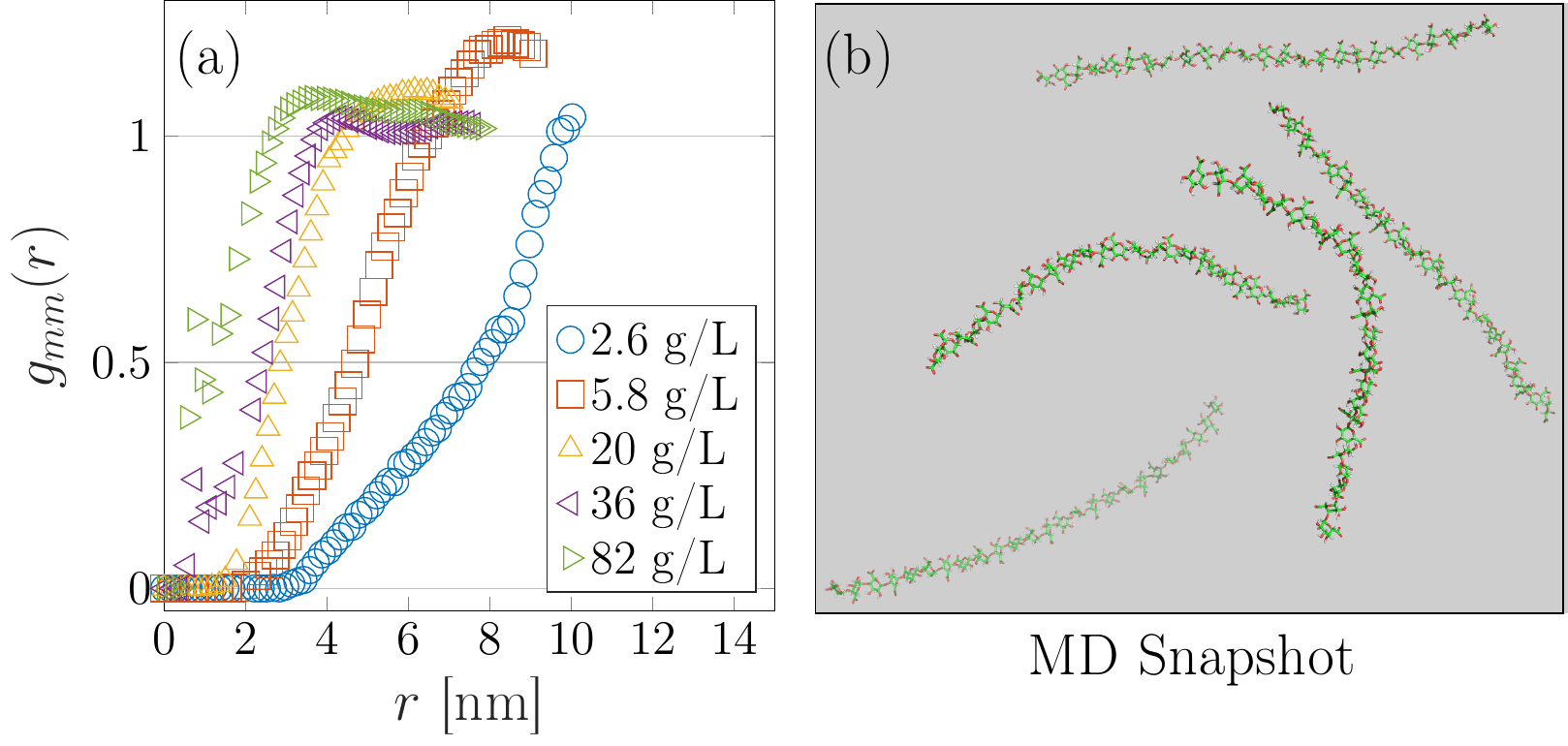}
	\caption{(a) showing chain-chain radial distribution functions $g_{mm}(r)$ calculated from MD simulations with 5 DP25 chains at concentrations of 2.60, 5.80, 20.00, 36.00, g/L. (b) a snapshot of one point during the simulation.}
	\label{fig:gr_dp25}
\end{figure}

\begin{figure}
	\includegraphics[width=0.9\columnwidth]{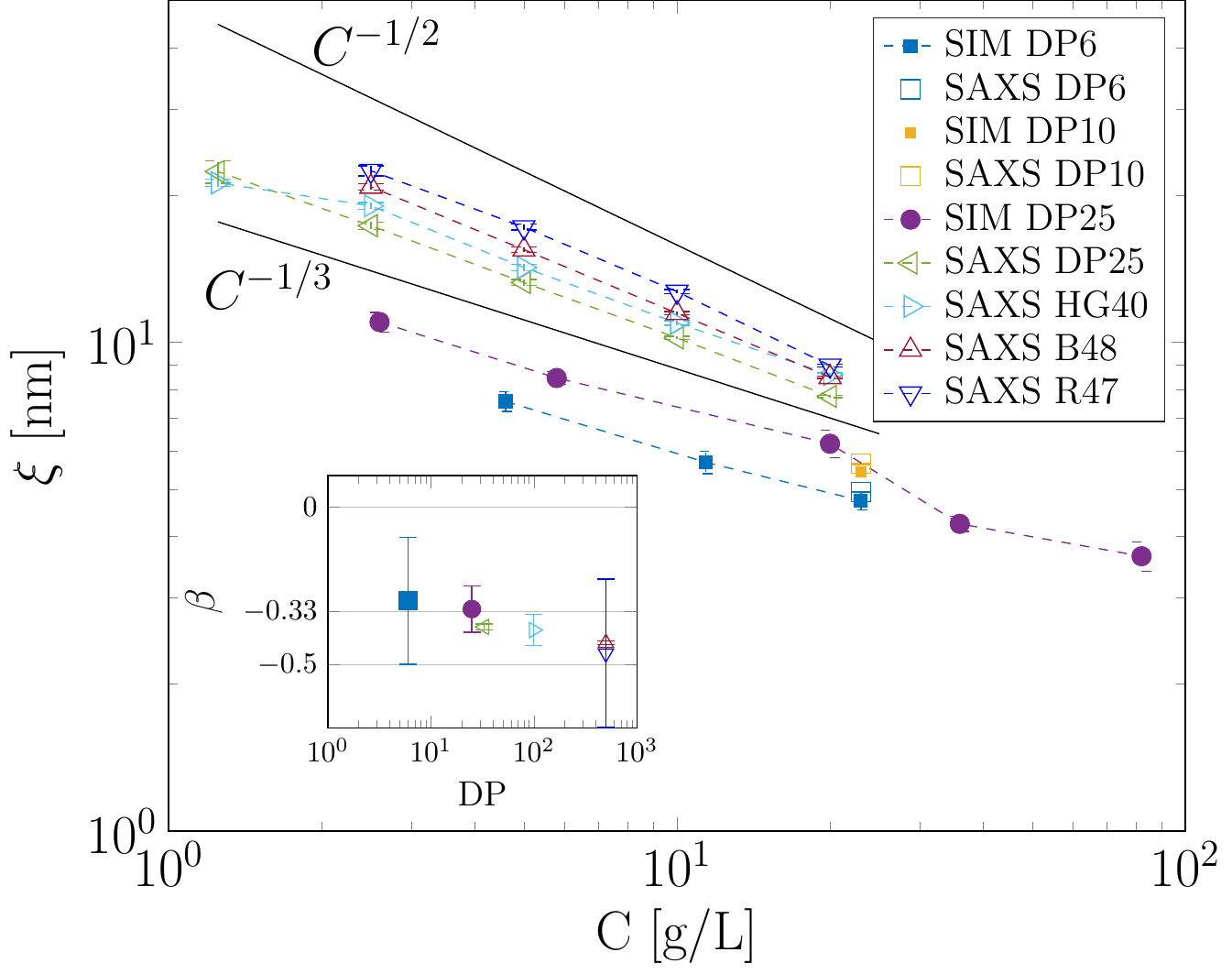}
	\caption{Correlation lengths, $\xi$, extracted from both SAXS and MD simulations as a function of concentration. The insert shows extracted exponents from power law fits as a function of degree of polymerization (DP). Error bars represent 95 \% confidence intervals.}
	\label{fig:6_zeta_conc}
\end{figure}

The effects of concentration on the inter-particle effects were further investigated for a range of different degrees of polymerization using both simulation and SAXS. Here we expect the oligomeric samples to display $\xi \propto C^{1/3}$ scaling if their solution structure can be approximated by points in space \cite{barrat1996theory,de1976remarks}. The polymeric samples would be expected to produce $\xi \propto C^{1/2}$ scaling reflecting the space filling of the mesh-like structure of solutions significantly far above the overlap concentration \cite{barrat1996theory,de1976remarks}. Figure \ref{fig:gr_dp25}(a) shows $g_{mm}(r)$ calculated for 5 $DP=25$ chains in simulation boxes corresponding to differing concentrations,(shown in the legend) over a simulation time of 20 ns. Figure \ref{fig:gr_dp25}(b) shows a snap-shot of the simulation where the individual chain conformations can be observed. Here we extract the correlation length both from SAXS experiments (using the peak in the intensity data) and simulation data. The results are shown in figure \ref{fig:6_zeta_conc} where the calculated correlation lengths at different concentrations for a variety of different degrees of polymerization (DP) are shown.

For the SAXS measurements the DP varies from 6 to approximately 500 (for both the blocky (B48) and random (R47) samples). For the samples with highest DP, namely B48 and R47, we see a power law scaling of the correlation length with concentration with an exponent approaching 0.5, (the insert shows power law fit exponents versus DP), consistent with the semi-dilute regime. At intermediate DP values a sample named HG40 with $DP \approx 100$ and a blocky charge pattern with a ratio of 60 \% of the chain having charged monomers have been investigated. This sample reveals a power-law between the two predicted values of 1/3 and 1/2. Interestingly, this is similar to the smaller DP sample labeled as DP25, which has DP distribution of 25 to 50, and $100\%$ unmethylesterified homogalacturonic acid. For the lowest DP samples, those with DP of 6, the scaling laws were only obtained at multiple concentrations using simulation, as limited amounts of material were available for experimental work. However, the SAXS results for the DP6 and DP10 samples at 23 g/L are included for comparison. For the DP6 sample the inter-particle effects were previously outlined and we see that at 23 g/L concentration the SAXS and simulation report similar $\xi$ values. The scaling for the DP6 sample obtained using simulation is $-0.3 \pm 0.2$ consistent with the predicted value of $1/3$, (although the errors are large as only three concentrations were simulated, due to limited computational resources being available). The multi-chain MD simulations show the DP25 sample scales with a power law of $-0.32 \pm 0.07$ consistent with  the predicted value for polyelectrolyte solutions below the overlap concentration. The simulation data indeed captures the scaling laws well and agrees well with the SAXS data where available. While the SAXS measurements of the DP25 sample show a significant difference from the simulation in terms of absolute values, the sample obtained on which the experiment was carried out had a broad distribution of DP (25-50 residues) so that an absolute comparison with the simulation of 25mers might be expected to show some differences.

\section{\label{sec:conclusion}Conclusion}
We have presented the results of SAXS experiments and multi-chain MD simulations investigating the solution conformations of homogalacturonans as model semiflexible polyelectrolytes. Comparisons between the scattering obtained from isolated chains and the output of calculations based on available atomic coordinates agree well. 

The persistence length was found to not depend directly on the charge density of the chains but instead was largely due to the intrinsic flexibility of the chain backbone (which is dependent on the helical conformation of the chains). Furthermore, the chains were found to not display worm-like chain statistics. We attribute this to the flexibility being dependent on the  conformation of the monomers which can locally vary to produce flexible and stiffer regions.

For oligomers of galacturonic acid with DP values of 6 and 10, under conditions where the chains were interacting, it was found that muti-chain MD indeed captures the emergence of the "polyelectrolyte peak". Furthermore, the scaling laws of the correlation length with concentration were investigated by SAXS and MD and it was found that the expected scaling laws could be accurately reproduced, within the errors, for both techniques. 

\section{ACKNOWLEDGMENTS}
Nigel Kirby and Jessie Owens are thanked for assistance during SAXS measurements and sample preparation. Davide Mercadante is thanked for assistance with MD simulations. The NZ synchrotron group for travel funding. Part of this research was undertaken on the SAXS / WAXS beamline at the Australian Synchrotron, part of ANSTO. This work benefited from the use of the SasView application, originally developed under NSF Award DMR-0520547. SasView also contains code developed with funding from the EU Horizon 2020 programme under the SINE2020 project Grant No 654000. New Zealand eScience Infrastructure (NeSI) are acknowledged for help with the provision of supercomputing facilities and support. Post doctoral funding for BM was provided by the Ministry of Science and Technology, Taiwan under grant No. MOST 105-2221-E-007-137-MY3.

%

%
%
%
\section{Authors contributions}
SAXS experiments were performed by BM, TR, MW, and DM, the corresponding data was analyzed by BM. Molecular dynamics simulations were run and analyzed by AI. The manuscript was prepared by BM, MW, AI and HLC. All authors made an intellectual contribution. 

%
\bibliographystyle{epj}

%
%
%

\end{document}